\documentclass[twocolumn,preprintnumbers,amsmath,amssymb,prb]{revtex4}

\usepackage{graphicx}
\usepackage{dcolumn}
\usepackage{bm}
\usepackage{amssymb}
\usepackage{epstopdf}
\usepackage{color}
\usepackage{dsfont}
\usepackage{float}

\begin{document}

\title{Periodic array of quantum rings strongly coupled to circularly polarized light as a
topological insulator}

\author{V. K. Kozin$^{1,2}$}
\author{I. V. Iorsh$^1$}
\author{O. V. Kibis$^{3,2}$}\email{Oleg.Kibis(c)nstu.ru}
\author{I. A. Shelykh$^{1,2}$}
\affiliation{$^1$ITMO University, Saint Petersburg 197101, Russia}
\affiliation{$^2$Science Institute, University of Iceland, Dunhagi
3, IS-107, Reykjavik, Iceland}\affiliation{${^3}$Department of
Applied and Theoretical Physics, Novosibirsk State Technical
University, Karl Marx Avenue 20, Novosibirsk 630073, Russia}
%\date{\today}

\begin{abstract}
We demonstrate theoretically that a strong high-frequency
circularly polarized electromagnetic field can turn a
two-dimensional periodic array of interconnected quantum rings
into a topological insulator. The elaborated approach is
applicable to calculate and analyze the electron energy spectrum
of the array, the energy spectrum of the edge states and the
corresponding electronic densities. As a result, the present
theory paves the way to optical control of the topological phases
in ring-based mesoscopic structures.
\end{abstract}

\maketitle

\section{Introduction}
Symmetries play crucial role in modern science since they
determine physical properties of various systems. Particularly,
the translational and inversion symmetries together with the
time-reversal symmetry define the electronic structure of solids.
If one of the symmetries is broken, the electron energy spectrum
complicates and the electron system can reach topologically
nontrivial phases~\cite{TopolQual}. So, the breaking of inversion
symmetry in semiconductor structures can lead to the transition
from the normal semiconducting state to the topological insulator
--- the matter that behaves as an insulator in its interior but
whose edges contain conducting electronic
states~\cite{TIReview,Qi_2011,Bernevig_book}. The breaking of
time-reversal symmetry --- for instance, by application of a
magnetic field --- radically changes the electron energy spectrum
as well, resulting in the discrete set of Landau levels within the
bulk and the chiral edge states at the boundaries. In turn, the
Coulomb interaction can further complicate the physical picture
leading to the incompressible Laughlin states and other exotic
phases of matter~\cite{QHEReview}.

It has been recently shown that topologically nontrivial phases
may arise not only in the condensed-matter electronic systems but
also in photonic structures~\cite{Haldane2008, Khanikaev2013} or
in the strongly coupled light-matter systems based on cavity
polaritons \cite{Karzig-PRX-2015, Bardyn-PRB-2015,
Nalitov-Z,KagomePolariton}. Moreover, in ring-like nanostructures
(particularly, in mesoscopic ballistic rings) the coherent
coupling of electrons to a circularly polarized electromagnetic
field leads to the appearance of an effective $U(1)$ gauge field
which breaks the time-reversal symmetry. Particularly, this field
results in the physical nonequivalence of clockwise and
counterclockwise electron rotations in an irradiated QR similarly
to a stationary magnetic
field~\cite{KibisRing,HelgiRing,Kibis_2015}. The field-controlled
interference of the electron waves corresponding to these
rotations opens a way to an optical tuning of QR
arrays~\cite{MehediChain}. Currently, QRs are actively studied
both experimentally and theoretically as a basis for various
nanoelectronic applications~\cite{Fomin_book}. Developing this
scientific trend in the present article, we demonstrate
theoretically that a circularly polarized field can turn a
two-dimensional (2D) array of interconnected QRs into a
topological insulator. It should be noted that topological
properties of electrons strongly coupled to light are under
consideration currently for various condensed-matter systems,
including graphene, semiconductor structures, etc (see, e.g.,
Refs.~\onlinecite{Yao_2007,Oka_2009,Kitagawa_2010,Kitagawa_2011,Lindner_2011}).
Therefore, the subject of the present research fits well the
growing tendencies in condensed-matter physics.

The paper is organized as follows. In Sec. II, we elaborate the
theory describing the stationary electronic properties of an
irradiated array of QRs. In Sec. III, we derive the electron
energy spectrum of the array, calculate the spectrum of electronic
edge states and the corresponding electronic densities, and
analyze the found edge states within the formalism based on the
Chern numbers. The last section contains the acknowledgments.

\section{Model}
Let us consider the 2D periodic array of interconnected QRs which
are irradiated by a circularly polarized electromagnetic wave with
the electric field amplitude, $\widetilde{E}_0$, and the
frequency, $\omega$ (see Fig.~1). In what follows, the field
frequency, $\omega$, is assumed to satisfy the two conditions.
Firstly, the field frequency should be far from interband resonant
frequencies of electrons in the periodic array. Under this
off-resonant condition, one can neglect the interband absorption
of the field. Secondly, the field frequency should satisfy the
condition $\omega\tau\gg1$, where $\tau$ is the electron life time
restricted by intraband scattering processes. Under this
high-frequency condition, the intraband (Drude) absorption of the
field by electrons can also be neglected. Since the wave is both
off-resonant and high-frequency (``dressing field'' within the
conventional terminology of quantum optics), the absorption of the
field does not erode the effects under consideration (for more
details, see also the discussion in
Refs.~\onlinecite{Morina_15,Kibis_16}). Correspondingly, the
effects caused by the dressing field substantially differ from the
effects induced in QRs by absorption of light (see, e.g.,
Refs.~\onlinecite{Pershin_2005,Rasanen_2007,Matos_2005,Koshelev}).
Physically, the circularly polarized dressing field breaks the
time-reversal symmetry and, therefore, effects on the electron
system in the QR similarly to a stationary magnetic
field~\cite{KibisRing}. In particular, electronic properties of a
dressed QR can be described by the effective stationary
Hamiltonian,
\begin{equation}\label{Heff}
H_{\mathrm{eff}}=\frac{(\hat{p}_\phi-eA_{\mathrm{eff}})^2}{2m_e},
\end{equation}
where $\hat{p}_\phi$ is the operator of electron momentum in the
QR, and
\begin{equation}
A_{\mathrm{eff}}=\frac{eE_0^2}{2Rm_e\omega^3} \label{Aeff}
\end{equation}
is the artificial $U(1)$ gauge potential produced by the
interaction between electrons in the QR and circularly polarized
photons of the dressing field~\cite{HelgiRing}. It follows from
Eqs.~(\ref{Heff})--(\ref{Aeff}) that the effective vector
potential, $A_{\mathrm{eff}}$, is responsible for the optically
induced Aharonov-Bohm effect~\cite{HelgiRing,Kibis_2015} and
results in the phase shift of an electron wave travelling between
nearest quantum point contacts (QPCs),
\begin{equation}
\phi_0=\frac{e\pi RA_{\mathrm{eff}}}{2 \hbar}=\frac{ \pi
e^2\widetilde{E}_0^2}{4m_e\hbar \omega^3}. \label{eq:phi0}
\end{equation}
Within the conventional scattering matrix
approach~\cite{Butticker1984,AB_bal_WL_2005}, the amplitudes of
electronic waves propagating in the array, $A_\pm$, $B_\pm$,
$C_\pm$, $D_\pm$, $F_\pm$, $G_\pm$, $H_\pm$, $I_\pm$, satisfy the
following set of equations,
\begin{equation}
\left(
\begin{array}{c}
A_- e^{{-iqd}/{2}} \\
F_+ \\
I_+ \\
\end{array}
\right)=S \left(
\begin{array}{c}
A_+ e^{{i qd}/{2}} \\
F_- e^{i \left({\pi q R}/{2}-\phi _0\right)} \\
I_- e^{i \left({\pi q R}/{2}+\phi _0\right)} \\
\end{array}
\right), \label{eq:leftQPC}
\end{equation}
\begin{equation}
\left(
\begin{array}{c}
B_+ e^{-i qd/2} \\
F_- \\
G_- \\
\end{array}
\right)=S \left(
\begin{array}{c}
B_- e^{{iqd }/{2}} \\
F_+ e^{i \left({\pi q R}/{2}+\phi _0\right)} \\
G_+ e^{i \left({\pi q R}/{2}-\phi _0\right)} \\
\end{array}
\right), \label{eq:upQPC}
\end{equation}
\begin{equation}
\left(
\begin{array}{c}
C_- e^{-i qd/2} \\
G_+ \\
H_+ \\
\end{array}
\right)=S \left(
\begin{array}{c}
C_+ e^{{iqd }/{2}} \\
G_- e^{i \left({\pi q R}/{2}+\phi _0\right)} \\
H_- e^{i \left({\pi q R}/{2}-\phi _0\right)} \\
\end{array}
\right), \label{eq:rightQPC}
\end{equation}
\begin{equation}
\left(
\begin{array}{c}
D_+ e^{-i qd/2} \\
H_- \\
I_- \\
\end{array}
\right)=S \left(
\begin{array}{c}
D_- e^{{iqd }/{2}} \\
H_+ e^{i \left({\pi q R}/{2}+\phi _0\right)} \\
I_+ e^{i \left({\pi q R}/{2}-\phi _0\right)} \\
\end{array}
\right), \label{eq:downQPC}
\end{equation}
where the scattering matrix is
\begin{equation}
S=\left(
\begin{array}{ccc}
\sqrt{1-2\varepsilon^2}  & \varepsilon  & \varepsilon  \\
\varepsilon  & \frac{-(1+\sqrt{1-2\varepsilon^2})}{2} & \frac{(1-\sqrt{1-2\varepsilon^2})}{2} \\
\varepsilon  & \frac{(1-\sqrt{1-2\varepsilon^2})}{2} & \frac{-(1+\sqrt{1-2\varepsilon^2})}{2} \\
\end{array}
\right), \label{eq:S1}
\end{equation}
$\varepsilon$ is the electron transmission amplitude through QPCs
($0\leq\varepsilon\leq1/\sqrt{2}$), $\hbar q=\sqrt{2m_e E}$ is the
electron momentum, $m_e$ is the electron effective mass, $E$ is
the electron energy, and $\phi_0$ is the field-induced phase shift
(\ref{eq:phi0}).
\begin{figure}[h]
\includegraphics[width=1\linewidth]{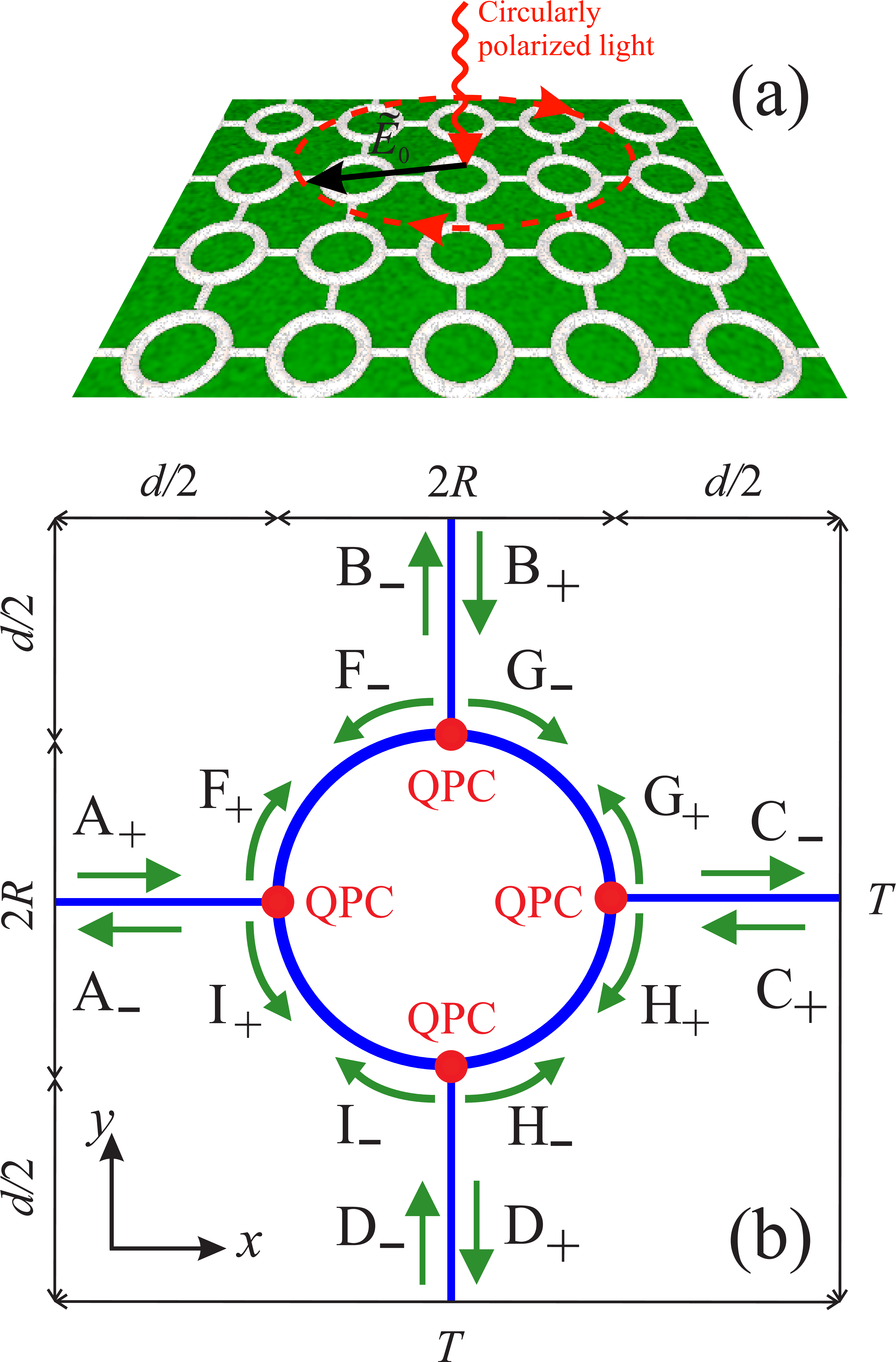}
\caption{Sketch of the system under consideration: (a) The 2D
periodic array of quantum rings (QRs) irradiated by a circularly
polarized electromagnetic wave with the electric field amplitude
$\mathbf{\widetilde{E}}_0$; (b) The elementary cell of the 2D
array with the period $T$, which consists of a QR with the radius
$R$, four leads with the length $d/2$ and four quantum point
contacts (QPC). The green arrows correspond to electron waves
traveling in different ways with the amplitudes $A_\pm$, $B_\pm$,
$C_\pm$, $D_\pm$, $F_\pm$, $G_\pm$, $H_\pm$, $I_\pm$.}
\label{ris:sketch}
\end{figure}
Applying the Bloch theorem to the considered periodic array of
QRs, we arrive at the equations
\begin{equation}
\left(
\begin{array}{c}
 C_- \\
 C_+ \\
\end{array}
\right)=e^{i k_x T}\left(
\begin{array}{c}
 A_+ \\
 A_- \\
\end{array}
\right) , \label{eq:Bloch_x}
\end{equation}
\begin{equation}
\left(
\begin{array}{c}
 B_+ \\
 B_- \\
\end{array}
\right)=e^{i k_y T}\left(
\begin{array}{c}
 D_- \\
 D_+ \\
\end{array}
\right) , \label{eq:Bloch_y}
\end{equation}
where $\mathbf{k}=(k_x,k_y)$ is the electron wave vector
originated from the periodicity of the array and $T=d+2R$ is the
period of the array. Mathematically,
Eqs.~(\ref{eq:leftQPC})-(\ref{eq:downQPC}) and
Eqs.~(\ref{eq:Bloch_x})-(\ref{eq:Bloch_y}) form the homogeneous
system of sixteen linear algebraic equations for the sixteen
amplitudes $A_\pm$, $B_\pm$, $C_\pm$, $D_\pm$, $F_\pm$, $G_\pm$,
$H_\pm$, $I_\pm$. Solving the secular equation of this algebraic
system numerically, we arrive at the sought electron energy
spectrum of the irradiated array, $E(\mathbf{k})$, which is
discussed below.

\section{Results and discussion}
The first six energy bands, $E(\mathbf{k})$, of the 2D infinite
periodic array of QRs with transparent QPCs
($\varepsilon=1/\sqrt{2}$) are plotted in Figs.~2a--2c for various
directions $\Gamma$-X-M-$\Gamma$ in the Brillouin zone of the
array (see Fig.~2d) and various phase incursions, $\phi_0$. In the
absence of the dressing field ($\phi_0=0$), the electron energy
spectrum plotted in Fig~2a has no gaps in the density of
electronic states and consists of two flat bands ($1$ and $6$),
which correspond to electron states localized within QRs, and four
bands ($2$--$5$) corresponding to delocalized electron states
propagating along the array. The dressing field ($\phi_0\neq0$),
firstly, results in coupling between localized and delocalized
electron states (see the dispersions of the $6$-th band in
Figs.~2a and 2b) and, secondly, induces the energy gaps between
different bands (see the green strips in Figs.~2b--2c). In what
follows, the gaps between the bands with the numbers $i$ and $j$
are denoted as $\Delta_{ij}$. It should be noted that the values
of the field-induced gaps depend periodically on the dressing
field as it is shown in Fig.~2e. Let us demonstrate that edge
states
--- both topologically protected and unprotected --- appear within
these field-induced band gaps.

Following the conventional theory of topological
insulators~\cite{Bernevig_book}, we have to fold the Brillouin
zone pictured in Fig.~2d as a torus $T^2$. Then the Chern number
corresponding to the $n$-th band of the considered periodic array
is defined as
\begin{equation}\label{ChN}
C_n=\frac{1}{2\pi i}\int_{T^2}d^2 k F_{xy}(\mathbf{k}),
\end{equation}
where $F_{xy}(\mathbf{k})={\partial A_y}/{\partial k_x}-{\partial
A_x}/{\partial k_y}$ is the field strength associated with the
Berry connection, $A_j(\mathbf{k})=\langle
n(\mathbf{k})|\frac{\partial}{\partial k_j}|n(\mathbf{k})\rangle$
is the vector potential of the field, and $|n(\mathbf{k})\rangle$
is the normalized Bloch wave function of the $n$-th
band~\cite{Bernevig_book}. Applying Eq.~(\ref{ChN}) to the Bloch
functions found from Eqs.~(\ref{eq:leftQPC})--(\ref{eq:Bloch_y}),
we arrive at the Chern numbers of the considered bands: $C_1=0$,
$C_2=-1$, $C_3=1$, $C_4=1$ and $C_5=-1$. According to the
bulk-boundary correspondence~\cite{Bernevig_book}, the sum of the
Chern numbers of the bands below a certain gap is equal to the
number of topologically protected edgemodes in the gap (per each
boundary):
\begin{equation}\label{TS}
N=\Big|\sum_{i}C_i\Big|.
\end{equation}
\begin{figure}[h!]
\includegraphics[width=0.8\linewidth]{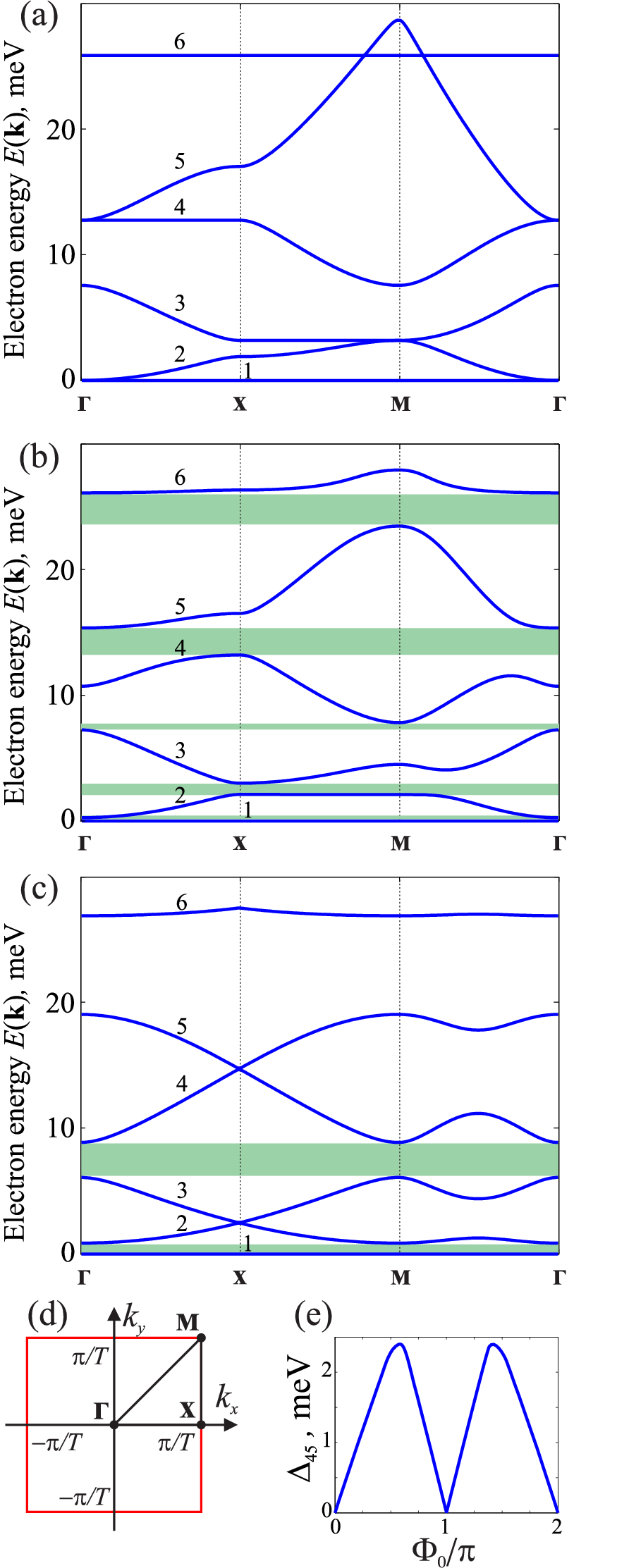}
\caption{Electron energy spectrum, $E(\mathbf{k})$, of the first
six bands ($1$--$6$) in the 2D infinite periodic array of
GaAs-based QRs with the radius $R=30$~nm, the period $T=100$~nm
and the electron transmission amplitude,
$\varepsilon=1\slash\sqrt{2}$, in the presence of a circularly
polarized electromagnetic field with the different phase
incursions, $\phi_0$: (a) $\phi_0=0$; (b) $\phi_0=\pi\slash9$; (c)
$\phi_0=\pi\slash4$. The green strips depict the field-induced
band gaps, the plot (d) shows the first Brillouin zone of the 2D
periodic array of QRs and the plot (e) presents the dependence of
the band gap , $\Delta_{45}$, on the field-induced phase incursion
along a whole QR, $\Phi_0=4\phi_0$.} \label{ris:3_graphs}
\end{figure}
\begin{figure}[h!]
\includegraphics[width=0.8\linewidth]{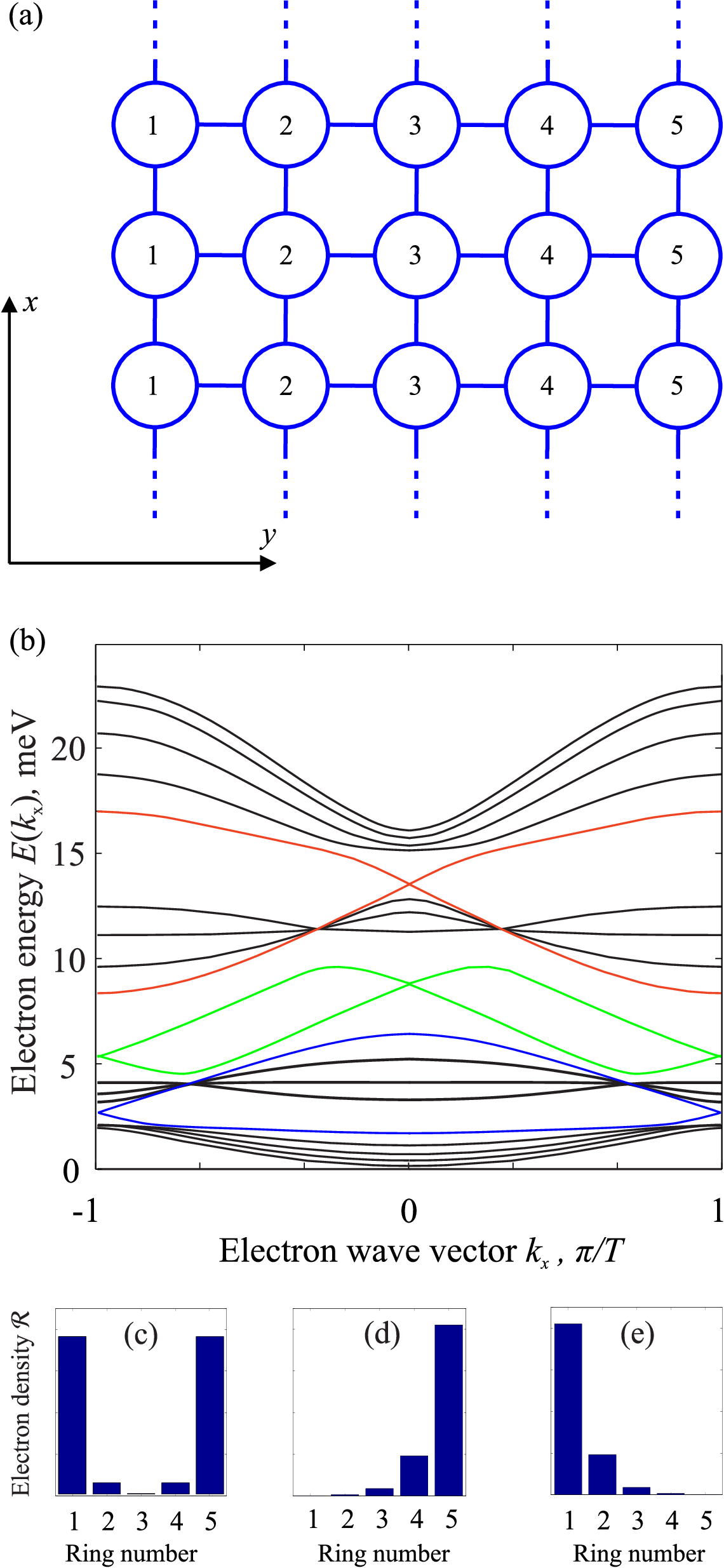}
\caption{Electronic properties of the finite array of QRs: (a)
Sketch of the finite array of QRs which is infinite along the $x$
axis and consists of five QRs along the $y$ axis; (b) The energy
bands, $E(\mathbf{k})$, of the finite array consisting of
GaAs-based QRs with the radius $R=30$~nm, the period $T=100$~nm
and the electron transmission amplitude,
$\varepsilon=1\slash\sqrt{2}$, in the presence of a circularly
polarized electromagnetic field with the phase incursions,
$\phi_0=\pi/9$. The red and blue lines correspond to the
topologically protected edge states, whereas the green lines
correspond to unprotected ones; (c)--(e) distribution of the
electron density, $\cal{R}$, in QRs with the different numbers for
the state corresponding to the intersection of the red branches at
$k=0$ in the plot (c), the state corresponding to the upper red
branch at $k_xT/\pi=-0.007$ in the plot (d), and the state
corresponding to the upper red branch at $k_xT/\pi=0.007$ in the
plot (d).} \label{ris:edge_states}
\end{figure}
It follows from Eq.~(\ref{TS}) that two branches of the
topologically protected edge states must exist within each of the
two non-zero band gaps, $\Delta_{23}$ and $\Delta_{45}$, if the
array of QRs is of finite size along one dimension.  To find
energy spectrum of the branches, let us restrict the consideration
by the array which is infinite along the $x$ axis and includes
only five QRs along the $y$ axis (see Fig.~3a). The energy
spectrum of electrons in this array can be easily calculated
within the approach developed in Sec.~II with the only difference:
The elementary cell of the finite array pictured in Fig.~3a
consists of five QRs, where the edge rings (with the numbers $1$
and $5$) have three QPCs unlike the others. The energy spectrum of
lowest bands of delocalized electrons in this array,
$E({\mathbf{k}})$, is plotted in Fig.~3b. As a first consequence
of finite size of the array along the $y$ axis, each delocalized
band of the 2D infinite array (see the branches $2$--$5$ pictured
in Figs.~2a--2c) is split into five subbands in Fig.~3b. As a
second consequence, the branches of edge states within the band
gaps appear (see the red, blue and green curves in Fig.~3b). The
edge character of these branches follows clearly from the
calculated electron density, $\cal{R}$, which has its maximum at
edge QRs with the numbers $1$ and $5$ (see Figs.~3c--3e). It
follows from the aforesaid that the branches of edge states
plotted in red and blue
--- which connects two adjacent bands
--- are topologically protected since their numbers satisfy the condition
(\ref{TS}). On the contrary, the edge states plotted in green are
not topologically protected since they correspond to the zero
number (\ref{TS}).

It should be noted that the band gap, $\Delta_{34}$, can also be
opened by varying the electron transmission amplitude through
QPCs, $\varepsilon$. Namely, $\Delta_{34}\neq0$, if the
transmission amplitude is $\varepsilon<1/\sqrt{2}$. However, this
gap opening does not result in topological edge states since the
corresponding number (\ref{TS}) is zero. As a consequence, only
field-induced band gaps can turn the periodic array of QRs into
topological insulator and, therefore, the strong electron coupling
to a circularly polarized dressing field is crucial for the effect
under consideration. Physically, this follows from the fact that
the circularly polarized dressing field effects on electron system
in a QR similarly to a stationary magnetic field (see
Eqs.~(\ref{Heff})--(\ref{eq:phi0}) and the detailed discussion in
Refs.~\cite{KibisRing,HelgiRing,Kibis_2015}). As a consequence,
formation of the edge electron states similar to those appearing
in the quantum Hall effect takes place in the considered periodic
array of irradiated QRs.

Finalizing the discussion, we have to formulate the conditions of
observability of the predicted effects. From this viewpoint, there
are the two fundamental restrictions: (i) the mean free path of
electron for inelastic scattering processes should be much greater
than the array period $T$; (ii) the time of electron traveling
through a ring should be much more than the field period,
$2\pi/\omega$. In the considered case of GaAs-based QRs with the
Fermi energy of meV scale and the array period $T\sim10^{-5}
\textnormal{cm}$, the both conditions can be satisfied for a
dressing field near the THz frequency range. Since the calculated
field-induced band gaps are of meV scale for such a dressing field
with the intensity $I\sim$~kW/cm$^2$, the discussed topological
edge states can be detected in state-of-the-art experiments. It
should be noted also that the present theory can be applied to
describe the same array in the presence of a stationary uniform
magnetic field directed perpendicularly to the array: One need
only to replace the phase shift (\ref{eq:phi0}) with the shift
arisen from the magnetic field.

In conclusion, we developed the theory of electronic properties of
the 2D periodic array of interconnected ballistic quantum rings
(QRs) interacting with an off-resonant circularly polarized
high-frequency electromagnetic wave (dressing field). It was
demonstrated that the Aharonov-Bohm effect induced by the dressing
field substantially modifies the electron energy spectrum of the
array, opening band gaps and producing edge states
--- both topologically protected and unprotected --- within the
field-induced gaps. As a result, the light-induced topological
insulator appears. The present theory paves the way to optical
control of the electronic properties of QR-based mesoscopic
structures.

\begin{acknowledgments}
The work was partially supported by RISE Program (project CoExAN),
FP7 ITN Program (project NOTEDEV), Russian Foundation for Basic
Research (Projects 16-02-01058 and 17-02-00053), Rannis Project
163082-051, and Ministry of Education and Science of Russian
Federation (Projects 3.4573.2017/6.7, 3.2614.2017/4.6,
3.1365.2017/4.6, 3.8884.2017/8.9 and 14.Y26.31.0015).
\end{acknowledgments}

\end{document}